# Bio-resorbable magnetic tunnel junctions

*Dong-Jun Kim[1†], Beom Jin Kim[2†], Hee-Chang Shin[2], Jeongkeun Kim[2], Yuchen Pu[1], Shuhan Yang[1], Xinhou Chen[1], Byong-Guk Park[3], Jong-Hyun Ahn[2]\*, and Hyunsoo Yang[1]\**

[1]Department of Electrical and Computer Engineering, National University of Singapore, Singapore

[2]School of Electrical and Electronic Engineering, Yonsei University, Seoul, Republic of Korea

[3]Department of Materials Science and Engineering, KAIST, Daejeon, Republic of Korea

[†] These authors equally contributed to this work.
*E-mail: eleyang@nus.edu.sg, ahnj@yonsei.ac.kr

**Magnetic tunnel junctions (MTJs) play a crucial role in spintronic applications, particularly data storage and sensors. Especially as a non-volatile memory, MTJs has received substantial attention due to its CMOS compatibility, low power consumption, fast switching speed, and high endurance. In parallel, bio-resorbable electronics have emerged as a promising solution for systems requiring temporary operation and secure data disposal, especially in military, intelligence, and biomedical systems where devices must safely disintegrate under physiological conditions. In this study, we investigate the bio-resorbability of MTJ by analyzing the dissolution behavior of its nanometer-thick constituent layers in phosphate-buffered saline (PBS) solution at pH 7.4, simulating physiological environments. The MTJ structures, composed of bio-resorbable materials, exhibit well-controlled degradation behaviors. Critically, as one of the ferromagnetic layers dissolves, binary information is irreversibly lost, within 10 hours of immersion. These findings highlight the potential of MTJs not only as high-performance memory**

**elements but also as secure, transient data storage platforms. The ability to modify the dissolution lifetime by materials and thickness selection offers unique advantages for short-lived implantable devices, paving the way for integrating spintronic functionality into next-generation bioresorbable electronics.**

## 1. Introduction

The magnetic tunnel junction (MTJ)[1-7], which exhibits tunneling magnetoresistance (TMR)[8], has recently attracted significant interest as a fundamental component in magnetoresistive random access memory (MRAM)[7, 9-10], magnetic sensor applications[10], and next-generation computing system[11-25], such as neuromorphic computing[11-14, 19, 22-23], process-in-memory[11, 15], and Ising computing[16, 20-21, 25], as well as in security technologies, including physical unclonable functions[17-18, 24]. TMR is particularly valued for its advantages in memory and computing applications, including low power consumption[26], fast switching speeds[6], non-volatility[9], high reliability[4, 24], and seamless integration with silicon-based semiconductor technology[27]. Structurally, an MTJ consists of nanometer-thick magnetic and non-magnetic multilayers, where information is stored based on the relative alignment of magnetizations between two ferromagnetic (FM) layers separated by a thin insulating barrier. The resistance is minimized when the magnetizations are parallel and maximized when antiparallel. Due to their binary switching capability, MTJs can function as both data storage and logic computing elements. As a result, MTJs have emerged as a promising next-generation memory technology.

In military and intelligence secure applications[28-29], where safeguarding sensitive information is critical, ensuring secure computing and irreversible data elimination is of paramount importance. In such contexts, bio-resorbable electronics[30-33], which can safely dissolve within the human body, offer a compelling approach to transient data storage. To prevent unauthorized access, it is crucial to ensure the irreversible elimination of stored data and associated computing components when necessary. These devices are already widely used in biomedical applications[34-37], including disposable implantable electronics[33, 38-39] and smart biosensors[32, 40]. Although their integration into security-focused technologies is still emerging, bioresorbable memory holds strong potential for enabling temporary data storage and automatic data deletion under physiological conditions.

In this context, we investigate the potential of MRAM as a bio-resorbable memory solution. MTJs, composed of nanometer-thick ultrathin magnetic and non-magnetic layers, offer a viable alternative to conventional non-volatile memory technologies that typically rely on bulkier structures. Because data storage in MTJs is determined solely by the magnetization alignment of two FM layers across a tunnel barrier, secure data erasure can be achieved simply by dissolving one of the FM layers. In addition, MTJs support a wide range of substrate materials, from flexible polymers[4] to rigid Si substrates[3], since their function depends only on the intrinsic material properties of the magnetic materials. By optimizing material selection for bio-resorbability, we aim to develop MTJ-based memory devices capable of transient operation and complete degradation under physiological conditions. We define the minimum structural requirements for bio-resorbable MTJs and quantify the dissolution behavior of individual constituent layers to evaluate their feasibility for secure, transient data storage applications.

## 2. Results and Discussion

### 2.1. Bio-resorbable MTJ concept and device structure

Figure 1a schematically illustrates the approach used to evaluate the bio-resorbable properties of the MTJ devices. To simulate the physiological environment of the human body[41], a 0.1 M phosphate-buffered saline (PBS) solution with a pH of 7.4 was prepared. The bio-resorbability of the MTJ devices was assessed by immersing both individual constituent materials and the complete MTJ multilayer stack in the PBS solution. Figure 1b presents the representative film stack of the bio-resorbable MTJ device, which was fabricated on either thermally oxidized silicon or glass substrates, depending on the experiments. The stack consists of a bottom tungsten (W) electrode, followed by the MTJ layer stack (CoFeB/MgO/CoFeB), and capped with a top W layer. MTJ devices are typically categorized into two types based on the relative magnetization alignment of the CoFeB layers[3]: out-of-plane MTJs (OOP-MTJs) and in-plane MTJs (IP-MTJs). In this study, we employed the OOP-MTJ configuration, which is the dominant structure in the current MRAM technology due to its superior high-density integration capability.

To improve the thermal stability of OOP-MTJs, an additional synthetic antiferromagnet (SAF) structure[42], composed of ultra-thin Co/Pt multilayers[43], was incorporated adjacent to the MTJ stack. The SAF layer stabilizes the magnetization of one CoFeB layer via the Ruderman–Kittel–Kasuya–Yosida (RKKY) coupling[44], enabling a clear distinction between the fixed and free magnetic layers in the MTJ stack. However, as discussed later, this SAF structure has a negligible impact on bio-resorbability due to its sub-nanometer-scale thickness. In IP-MTJs, the stabilization of one FM layer is typically achieved by exchange bias with an adjacent antiferromagnetic (AFM) layer, also allowing differentiation between fixed and free magnetic layers. The bio-resorbability of such AFM materials of iridium manganese (IrMn) and iron manganese (FeMn) is presented in Figure S1. Both materials exhibit limited degradation in PBS solution, indicating poor suitability for transient device applications under physiological conditions. Both IrMn and FeMn exhibited rapid initial dissolution within the first hour of immersion. However, after this early stage, the degradation rate significantly slowed, and the films remained largely intact throughout the subsequent period. This non-linear dissolution behavior appears to be characteristic of Mn-containing compounds, as previously reported in the literature[45], where the formation of stable oxides or passivation layers is believed to inhibit further degradation. This expanded discussion provides valuable context regarding the limitations of commonly used AFM layers in bio-resorbable spintronic devices and may guide future material selection strategies for bio-resorbable applications.

The TMR effect arises from the relative alignment of the two FM layers. When the magnetizations are antiparallel, the resistance becomes high, corresponding to a binary "1" state. Conversely, when the magnetizations are parallel, the resistance becomes low, representing a binary "0" state, as illustrated in Figure 1b. A key advantage of MTJs for bio-resorbable applications is that data storage depends solely on the magnetization alignment between the two FM layers. For secure data erasure, the complete dissolution of the MTJ stack is ideal. However, as depicted in Figure 1c, dissolving even a single FM layer is sufficient to disrupt the binary state, thereby effectively erasing the stored data. This can be achieved by removing only a few nanometers of materials, specifically, the top W and CoFeB layers, making MTJs a significantly more compact and efficient bio-resorbable memory solution compared to other bulkier next-generation memory devices. Once the magnetization alignment is lost, data recovery becomes physically impossible.

To fabricate the bio-resorbable OOP-MTJ devices, the optimized thin film stack was deposited using ultrahigh vacuum magnetron sputtering. As illustrated in Figure 2a, the devices were patterned into a Hall-bar bottom electrode structure ($10 \times 100$ μm$^2$) and a pillar structure (5 μm diameter) via photolithography, subsequently followed by physical etching using ion milling. To electrically isolate and protect the MTJ structure, a 50-nm-thick silicon nitride (SiN$_x$) passivation layer was deposited on the sidewalls of the pillar structure immediately after ion milling, without breaking the vacuum environment. Subsequently, a 50-nm-thick copper (Cu) top electrode was deposited via magnetron sputtering to establish electrical contacts on the MTJ structure. The fabricated OOP-MTJ device, composed of bio-resorbable materials, as shown in Figure 2b, exhibits a clear TMR response, achieving a TMR ratio of 75%, as shown in Figure 2c.

**2.2. Dissolution behavior of individual MTJ materials in PBS solution**

Figure 3 presents the transmittance variations of key MTJ materials, along with transient optical microscopy images of the Hall-bar structure patterns during dissolution in PBS solution at 37 °C. To quantitatively assess the dissolution behavior, UV-visible spectroscopy was employed to monitor transmittance changes over time. Each MTJ material was deposited as a thin film on a transparent glass substrate to facilitate optical analysis. In addition to the primary MTJ components, a 50-nm-thick Cu top electrode was deposited at multiple positions: on the top W layer of the MTJ pillar structure, above the $Si_3N_4$ passivation layer, and over the bottom W electrode for electrical contacts, as illustrated in Figure 2a. Therefore, understanding the dissolution behavior of W, CoFeB, MgO, and Cu layers is critical for optimizing bio-resorbable MTJ applications. To enable quantitative comparison, thicker individual films were prepared and tested.

The 10-nm W thin film exhibits rapid dissolution, with transmittance increasing from ~30% to over 96% across the measured wavelength range after 180 minutes in PBS solution (Figure 3a). In contrast, the 5-nm CoFeB and MgO thin films dissolve much more slowly. CoFeB serves as the magnetic layer responsible for data storage, while MgO functions as the tunnel barrier, both are essential components of MTJs. However, these layers are typically thinner than 5 nm in actual devices. The FM layer thickness depends on the MTJ type and target application. For example, OOP-MTJs[5, 46] designed for high-density memory typically use ~1 nm-thick FM layers, while IP-MTJs[7] for ultrafast computing may employ FM layers exceeding 2 nm. Therefore, the dissolution properties of FM layers can be optimized and tailored to specific application requirements. One important aspect to consider in the design of bio-resorbable MTJs is the post-deposition annealing process, which is commonly used to enhance the TMR ratio in CoFeB-based structures. Annealing typically induces boron (B) diffusion out of the CoFeB layer and leads to the formation of a crystallized CoFe phase. To examine the impact of this transformation on bio-resorbability, we investigated the dissolution of a 5-nm CoFe film in PBS solution (Figure S2). The CoFe film dissolved almost completely after just 30 minutes of immersion, indicating a much faster degradation rate than CoFeB. These results suggest that annealing-induced phase changes can accelerate degradation and should be carefully considered when designing bio-resorbable MTJ devices.

For enhanced bio-resorbability, alternative FM materials such as permalloy (NiFe) and cobalt (Co) were also explored (see Figure S3). Compared to CoFeB shown in Figure 3, both NiFe and Co exhibit significantly faster degradation properties, highlighting their potential as promising FM candidates for transient MTJ applications. Similarly, a thin MgO tunnel barrier (1-2 nm) is essential for maintaining tunneling characteristics in both MTJ types. Although a 5-nm-thick MgO film requires approximately five days for its transmittance to increase from ~92% to ~96% (Figure 3c), thinner MgO layers (1-2 nm) can dissolve more rapidly in PBS solution.

The top FM layer of the MTJ stack is protected by a W(5 nm)/Cu (50 nm) bilayer (Figure 2a). The 50-nm Cu top electrode plays a critical role in data protection in PBS solution and must be dissolved to expose underlying layers. Initially, the Cu film exhibits a transmittance of ~5% (Figure 3d). After five hours of immersion, transmittance increases to ~90%, indicating substantial degradation. These results confirm that both W and Cu layers dissolve efficiently in PBS solution, enabling subsequent exposure and dissolution of internal MTJ layers. In contrast to the bio-resorbable materials presented in Figure 3, platinum (Pt) and tantalum (Ta), which are commonly used heavy materials in various spintronic devices, exhibit poor degradation behavior in PBS solution because Pt, a noble metal, inherently resists oxidation and corrosion, and Ta forms a stable and protective tantalum pentoxide layer on its surface, which effectively serves as an oxidation barrier. This is confirmed by additional UV-visible spectroscopy results provided in Figure S3 for various constituent materials. The dissolution behavior is further corroborated by the transient optical microscopy images of the representative Hall-bar pattern on Si substrates (Figure 3e-h). Similar degradation trends were also confirmed through surface roughness changes of various films, as observed in atomic force microscopy images presented in Figure S4. After immersion, the surfaces become smoother and more clear, reflecting material-dependent dissolution behavior.

## 2.3. Dissolution behavior of the complete MTJ stack

The dissolution behavior of the complete MTJ stack, including an additional 10-nm SAF structure composed of ultrathin Co/Pt multilayers for differentiating fixed and free magnetic layers, was investigated at 37 °C (Figure 4; see Figure S5 for details on the SAF structure). Consistent with the dissolution behavior of individual materials presented in Figure 3, transmittance measurements obtained via UV-visible spectroscopy, along with optical microscopy images of the representative Hall-bar pattern, confirm the effective degradation of the bio-resorbable OOP-MTJ device. To facilitate quantitative analysis of the dissolution behavior, a modified stack configuration with a thicker bottom electrode was employed. The original 4-nm W bottom electrode (Figure 2a) was replaced with a W (10 nm)/ Ru (5 nm)/W (4 nm) tri-layer structure, which is commonly used in MRAM devices. This adjustment increased the total MTJ stack thickness to ~42 nm. Ruthenium (Ru) also serves as a promising candidate for bio-resorbable bottom electrodes, owing to its rapid dissolution behavior in PBS solution, as demonstrated in Figure S3.

Transmittance measurements indicate substantial degradation of the complete MTJ stack over 18 days, with transmittance increasing from ~12% to ~93% (Figure 4b). To further evaluate this process, transient longitudinal resistance measurements were performed on the representative Hall-bar pattern. The initial resistance of ~320 Ω increased to ~7.5 kΩ after 4 hours and exceeded measurable limits after 5 days of dissolution (Figure 4c). Additional insights into the dissolution were obtained through step height measurements (Figure 4d). Samples were fabricated on transparent glass substrates with distinct regions, one without and one with the MTJ stack, using the same representative Hall-bar pattern. The MTJ stack thickness decreased from ~42 nm to ~24 nm after just 4 hours, with the dissolution process reaching near saturation after 5 days, consistent with the transmittance data. To achieve data erasure, the device was exposed to a PBS solution for approximately four hours, resulting in an ~18 nm reduction in the step height. This change confirms that the upper critical 16-nm portion, comprising a 5 nm tungsten (W) top layer, a 9.6 nm SAF structure, and a 1.4 nm fixed CoFeB layer (see Figure S5 for the SAF structure), was completely dissolved. These results highlight the strong bio-resorbable potential of the MTJ structure. In cases where a 50-nm Cu top electrode is also present for MTJ devices in Figure 2a, approximately 10 hours of immersion is sufficient for data removal. The thicknesses of both the Cu top electrode and the SAF structure can be optimized, allowing for control over the total dissolution time, which can be shortened to just a few hours.

Interestingly, functional data loss can occur earlier than complete physical degradation in MgO-based MTJ devices due to the hygroscopic nature of the MgO tunnel barrier[47]. To investigate this possibility, we monitored the TMR behavior during the early stages of immersion in PBS solution. Devices with different initial TMR ratios were immersed for 1 to 5 hours before completely removing the Cu top electrode, and their TMR characteristics were subsequently measured (Figure S6). Notably, even after just 1 hour of immersion, the junction resistance dropped significantly, and TMR signals were no longer detectable. This rapid electrical failure highlights the critical role of the MgO barrier in determining the functional lifetime of bio-resorbable MTJs. These findings provide a lower bound for operational durability and suggest that alternative insulating materials may be needed to extend device lifetime under physiological conditions.

Furthermore, this process can be accelerated under physiological conditions, where elevated temperature and continuous mechanical agitation from bodily motion and internal processes enhance dissolution kinetics. Although pH levels vary across different internal organs, this study focuses on PBS solution at pH 7.4 as a representative physiological condition, providing a baseline understanding of dissolution behavior. While some constituent materials of MTJs, such as Pt, Co, and Ru, may raise concerns regarding long-term biocompatibility, this challenge can be mitigated through material substitution and compositional tuning of alloy systems. The intrinsic flexibility of the bio-resorbable MTJ platform allows for the integration of alternative ferromagnetic materials, such as Fe, MnGa, or NiFe. Fe, which is naturally present in the human body, can serve as biocompatible ferromagnetic components in bio-resorbable MTJs. In addition, certain alloys such as MnGa and NiFe may also be considered, provided that their compositions are carefully adjusted to minimize the release of potentially harmful ions. These materials could be applicable in specific scenarios where the dose, degradation rate, and ion clearance pathways can be carefully controlled, highlighting future opportunities for material selection strategies based on both functionality and biocompatibility. Overall, these findings confirm that MTJ-based bio-resorbable devices can effectively disintegrate in physiological environments, enabling secure and irreversible data erasure.

While conventional methods such as mechanical destruction, chemical dissolution, or even combustion are available for external data removal, *in vivo* degradation offers distinct advantages in biomedical applications where device retrieval is difficult or medically undesirable. In such cases, passive dissolution enables practical and automatic information loss without physical intervention, reinforcing the suitability of MTJs for secure, short-lived electronics. This establishes their feasibility not only for transient memory and computing systems but also for broader applications in bioelectronics and biomedical devices requiring built-in, on-demand data destruction. Depending on the target application, the operational lifetime and dissolution rate of bio-resorbable MTJ devices can be tuned by selecting appropriate constituent materials and adjusting the thickness of each layer.

## 3. Conclusion

This study demonstrates the potential of bio-resorbable MTJ devices as secure and transient memory platforms by investigating their dissolution behavior under physiologically relevant conditions. Experimental results show that a representative nanometer-thick magnetic multilayer used in MTJ devices can effectively degrade in PHS solution at pH 7.4, enabling irreversible data erasure within 10 hours through the controlled removal of key constituent layers. Since the data in MTJs are solely determined by the magnetization alignment of ultrathin FM layers separated by a thin insulating barrier, dissolving a single FM layer is sufficient to permanently eliminate stored information. This unique mechanism, combined with inherent advantages of MTJs, including fast switching speed, low power consumption, non-volatility, and seamless integration with CMOS platforms, makes them particularly well suited for secure memory applications requiring temporary operation and automatic data disposal. The dissolution kinetics can be tuned by adjusting layer thicknesses and further accelerated under physiological conditions such as elevated temperature or mechanical agitation. Overall, these findings confirm the feasibility of integrating MTJ-based memory into bio-resorbable electronic systems. Such devices offer a compelling combination of on-demand data erasure, transient functionality, and biocompatibility, paving the way for their use in secure, short-lived electronics. Moreover, the ability to fine tune the bio-resorbable lifetime by selecting appropriate materials and adjusting layer thickness provides a distinct advantage, particularly for expanding their applicability in implantable and next-generation biomedical systems.

## 4. Experimental section

*Sample preparation*: All thin-film samples were deposited using an ultra-high vacuum d.c. magnetron sputtering system (AJA ORION) on either thermally oxidized silicon or transparent glass substrates, depending on the experimental purpose. Hall-bar structures with a width of 10 μm and a length of 100 μm, serving as the bottom electrode of the MTJ and for quantitative dissolution analysis, were patterned by photolithography and etched using an ion milling system. A mask aligner (KarlSuss-MA6) was employed to align and pattern all structures. The MTJ pillar structure, with a diameter of 5 μm, was patterned at the center of the Hall-bar bottom electrode, followed by ion milling to define the dot shape. A 50-nm-thick $SiN_x$ passivation layer was subsequently deposited by r.f. sputtering after the etching step, without breaking the vacuum, to encapsulate the exposed sidewalls. Finally, a 50-nm-thick Cu contact electrode, patterned in a diagonal cross shape for electrical measurements, was fabricated using the same aligner and deposited via a physical vapor deposition system.

*TMR characteristics*: MR measurements were performed to evaluate the TMR characteristics of the representative bio-resorbable MTJ device. The setup consisted of a source meter (Keithley 2400) to supply a constant d.c. bias current and a nanovoltmeter (Keithley 2182A) to measure the voltage drop across the MTJ via a standard four-probe configuration. A bias current of 100 μA was applied from the bottom Cu electrode to the top Cu contact electrode to ensure stable current flow through the MTJ stack. During the measurement, an external magnetic field ($H_{ext}$) was swept perpendicularly to the film plane (along the z-axis) using an electromagnet to modulate the relative alignment of the FM layers in the MTJ. By recording the voltage response as a function of the applied magnetic field, the TMR ratio was extracted.

The resulting MR curve reveals the characteristic switching behavior of ~75% TMR ratio between parallel and antiparallel magnetization states, validating the functional integrity of the MTJ stack.

*Dissolution experiments*: To evaluate the bio-resorbability of the MTJ devices and their constituent materials, in vitro dissolution experiments were performed. Samples designated for dissolution testing were fully immersed in 0.1 M Phosphate-Buffered Saline (PBS; pH 7.4, Sigma-Aldrich) solution maintained at a physiological temperature of 37 °C in a constant-temperature incubator. The PBS solution was periodically refreshed to maintain stable pH and ionic strength throughout the experiment. Electrical resistance was measured using a four-probe configuration with a Keithley 2400 source meter and a Keithley 2182A nanovoltmeter. Surface topography was monitored using atomic force microscopy (Park Systems NX-10) operated in non-contact mode. Furthermore, changes in the optical properties of the thin films were tracked using ultraviolet–visible (UV-Vis) spectroscopy (Jasco V-650). Optical absorbance and/or transmittance spectra were measured in the 400–800 nm wavelength range at various time points during the dissolution process.

**Acknowledgements**

The research is supported by National Research Foundation (NRF) Singapore Investigatorship (NRFI06-2020-0015), Ministry of Education, Singapore, under Tier 2 (T2EP50123-0025), and Samsung Electronics Co., Ltd (IO241218-11518-01). J.H.A acknowledges support from the Ministry of Trade, Industry, and Energy (MOTIE) of Korea (20012355).

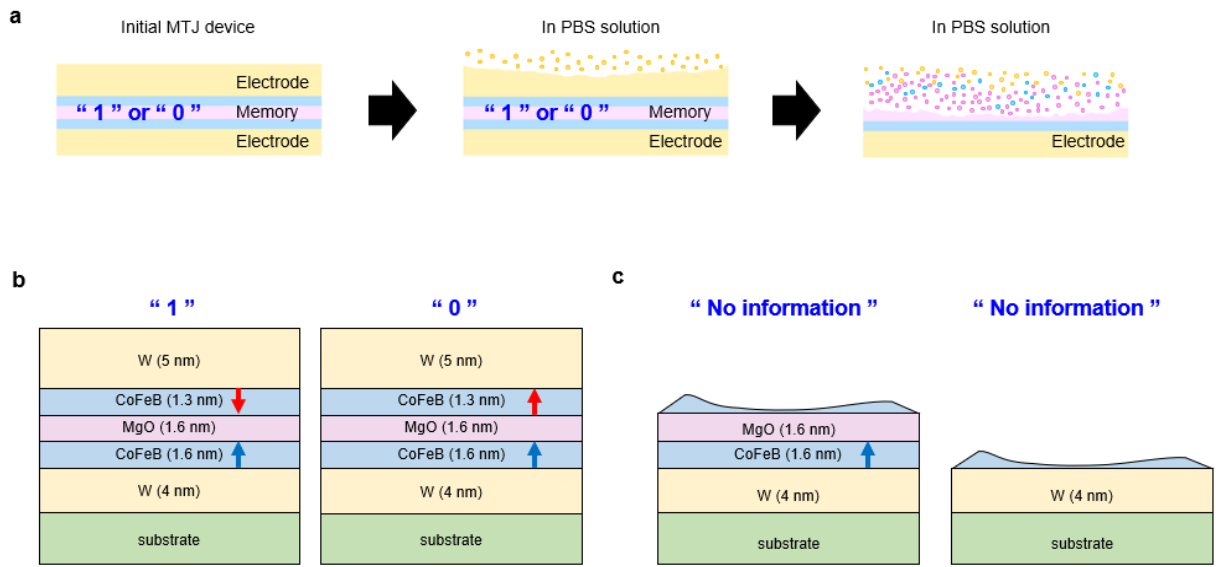

**Figure 1.** Bio-resorbable MTJ concept and device structure. a) Schematic illustration of the bio-resorbable memory concept using MTJs. Information stored through magnetization alignment can be irreversibly erased by dissolving magnetic layers. b) Binary state representation of MTJ resistance based on the magnetization alignment of CoFeB layers: low resistance (parallel alignment) as "0" and high resistance (antiparallel alignment) as "1". c) Conceptual diagram illustrating the data erasure process, in which dissolution of the top CoFeB and W layers disrupts the magnetization configuration, effectively eliminating stored information and preventing recovery.

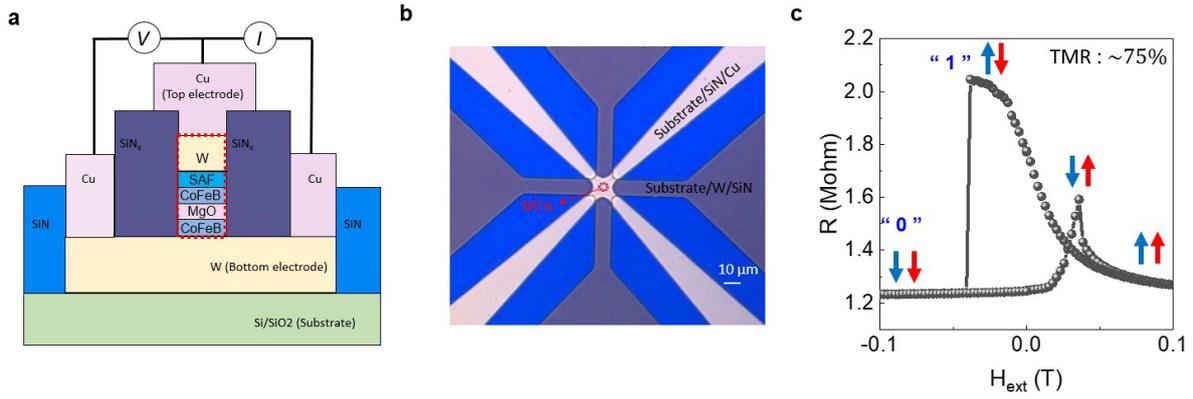

**Figure 2.** Fabrication and TMR characteristics of MTJ devices. a) Schematic illustration of the MTJ device, including the SiN$_x$ passivation layer and the Cu top electrode. b) Optical microscope image of the representative MTJ device, featuring an MTJ pillar with a diameter of 5 μm. c) External magnetic field (H$_{ext}$) driven TMR curve of the OOP-MTJ device, exhibiting a TMR ratio of 75% at room temperature, confirming reliable magnetic switching behavior. Red and blue arrows indicate the magnetization direction of each CoFeB layer.

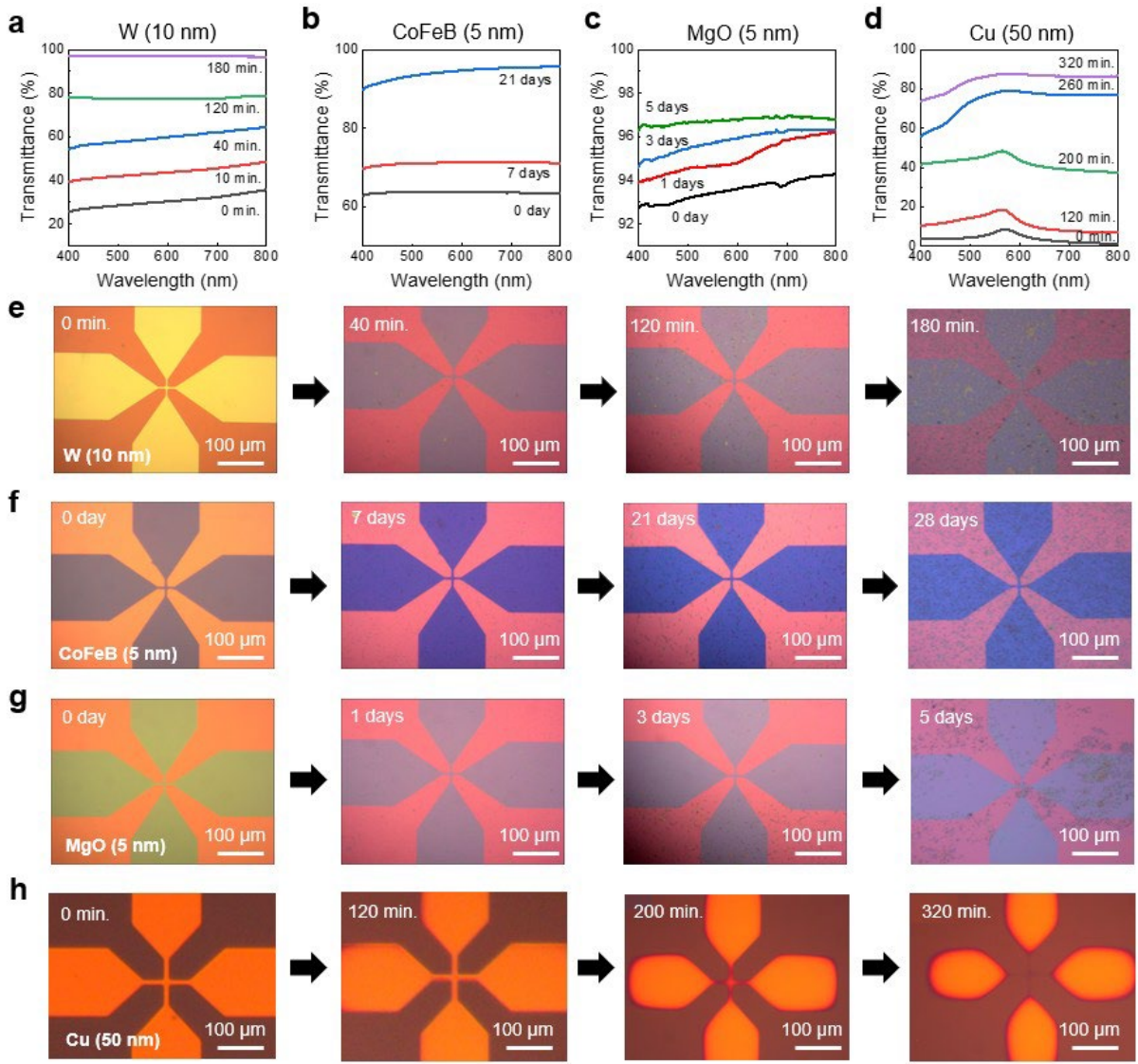

**Figure 3.** Dissolution behavior of individual MTJ materials in PBS solution at 37 °C. a) Transmittance change of a 10-nm-thick W film during dissolution in PBS (pH 7.4), showing rapid degradation over 180 minutes. b) Transmittance behavior of a 5-nm-thick CoFeB film under the same conditions, exhibiting a slower dissolution. c) Transmittance change of a 5-nm-thick MgO film, indicating gradual degradation over 5 days. d) Transmittance variation of a 50-nm-thick Cu film, showing substantial dissolution within 5 hours. e–h) Optical microscopy images of the Hall-bar pattern at different dissolution stages, visually confirming progressive, layer-by-layer degradation of the constituent materials.

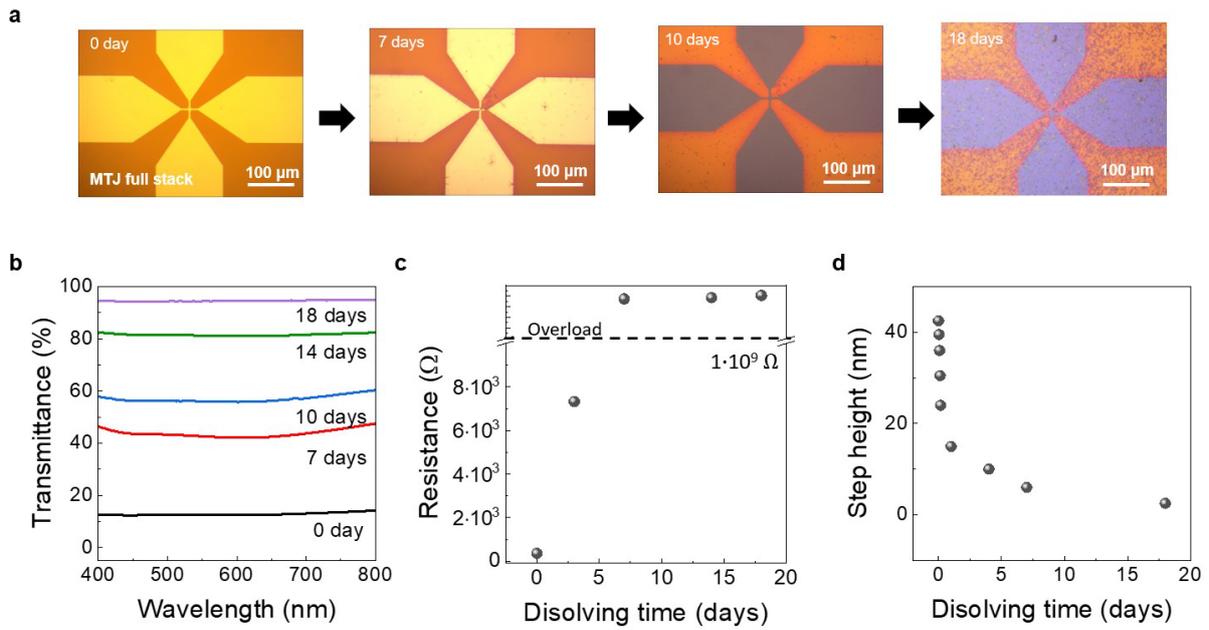

**Figure 4.** Dissolution kinetics of the complete MTJ stack including the SAF structure. a) Optical microscopy images of the Hall-bar pattern during the degradation of the complete MTJ stack, which includes a 10-nm SAF layer. b) Transmittance evolution of the MTJ stack in PBS solution (pH 7.4) over 18 days, increasing from ~12% to ~93%, indicating progressive dissolution. c) The resistance change of the Hall-bar pattern, showing degradation from ~320 Ω to beyond the measurable range after 5 days. d) Step height measurements of the MTJ stack during dissolution, decreasing from ~42 nm to ~24 nm within 4 hours, confirm the removal of the critical upper layers responsible for data storage.

# Bio-resorbable magnetic tunnel junctions

*Dong-Jun Kim[1†], Beom Jin Kim[2†], Hee-Chang Shin[2], Jeongkeun Kim[2], Yuchen Pu[1], Shuhan Yang[1], Xinhou Chen[1], Byong-Guk Park[3], Jong-Hyun Ahn[2]\*, and Hyunsoo Yang[1]\**

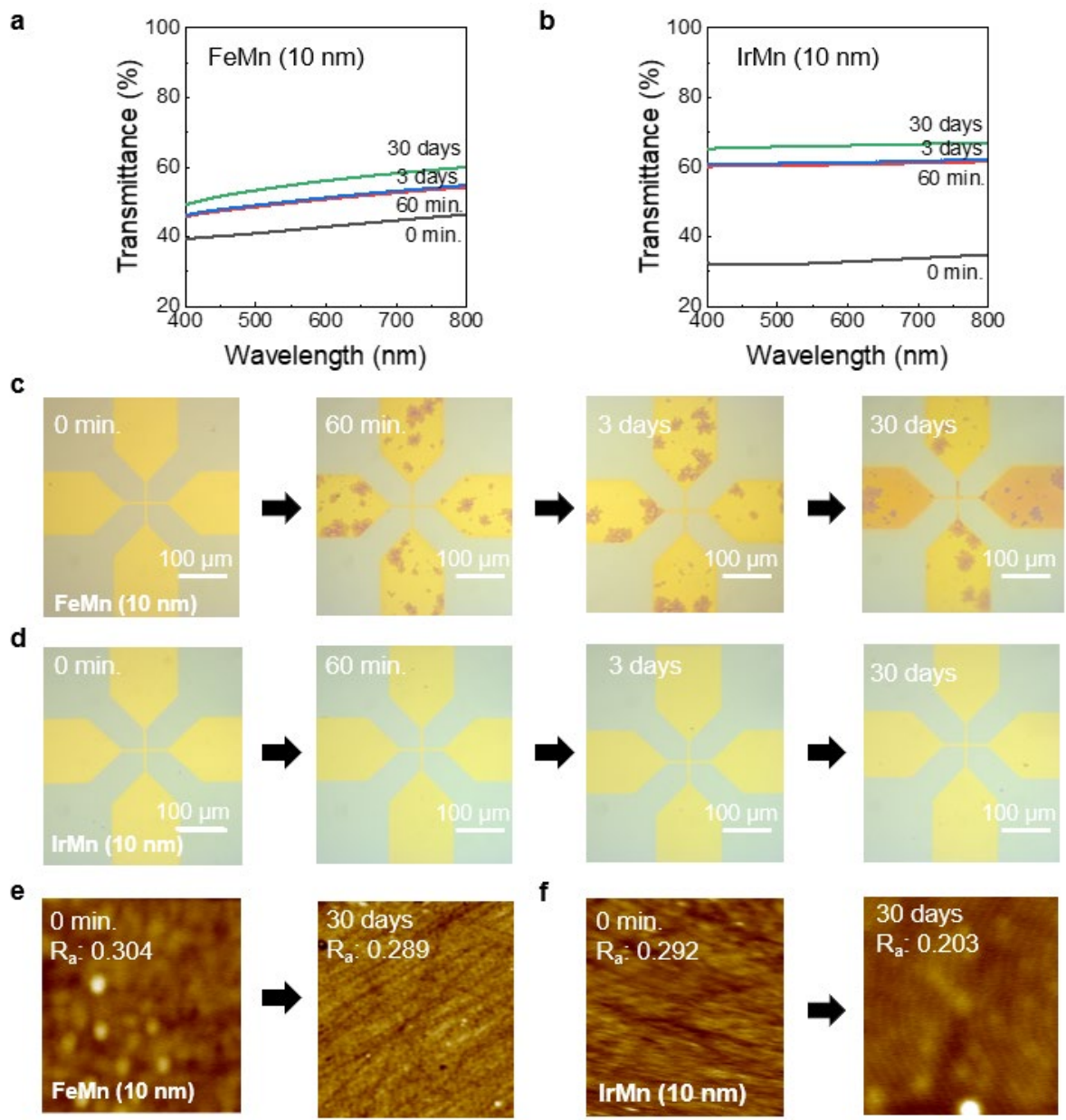

**Figure S1.** Dissolution behavior of individual antiferromagnetic (AFM) materials in PBS solution (pH 7.4) at 37 °C. a) Transmittance change of a 10-nm-thick $Fe_{50}Mn_{50}$ film during dissolution. b) Transmittance change of a 10-nm-thick IrMn film during dissolution. c,d) Corresponding optical microscopy images of the Hall-bar patterns at different dissolution stages of each AFM material. e,f) Corresponding atomic force microscopy images showing surface roughness evolution of each AFM material during dissolution. Both AFM materials exhibit more pronounced surface changes over 1 hour; however, visible residues remain, confirming poor bio-resorbability for both antiferromagnetic materials under physiological conditions.

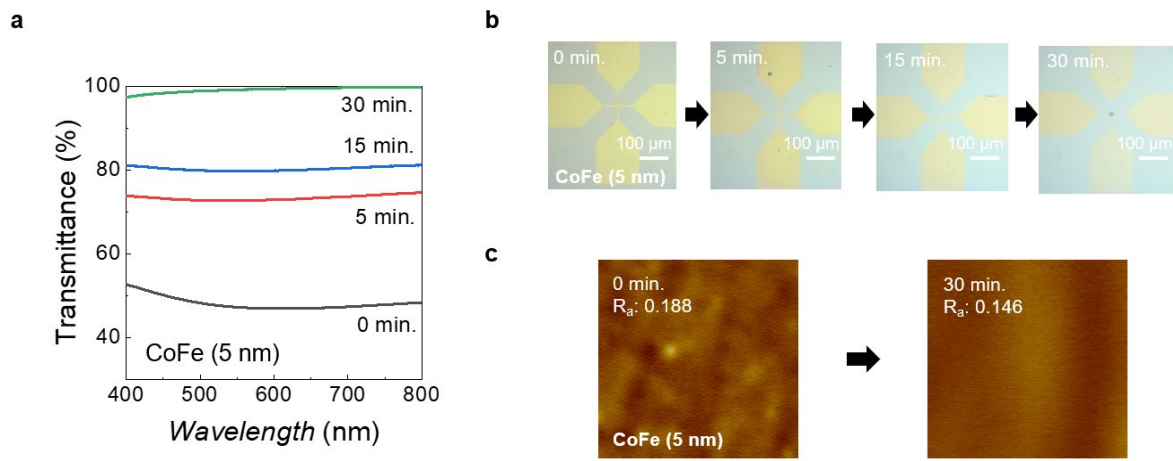

**Figure S2.** Dissolution behavior of a CoFe (5 nm) film in PBS solution at 37 °C. a) Transmittance change of a 5-nm-thick CoFe film during dissolution. b) Corresponding optical microscopy images of the Hall-bar patterns at different dissolution stages of the CoFe film. c) Corresponding atomic force microscopy images showing surface roughness evolution of the CoFe film.

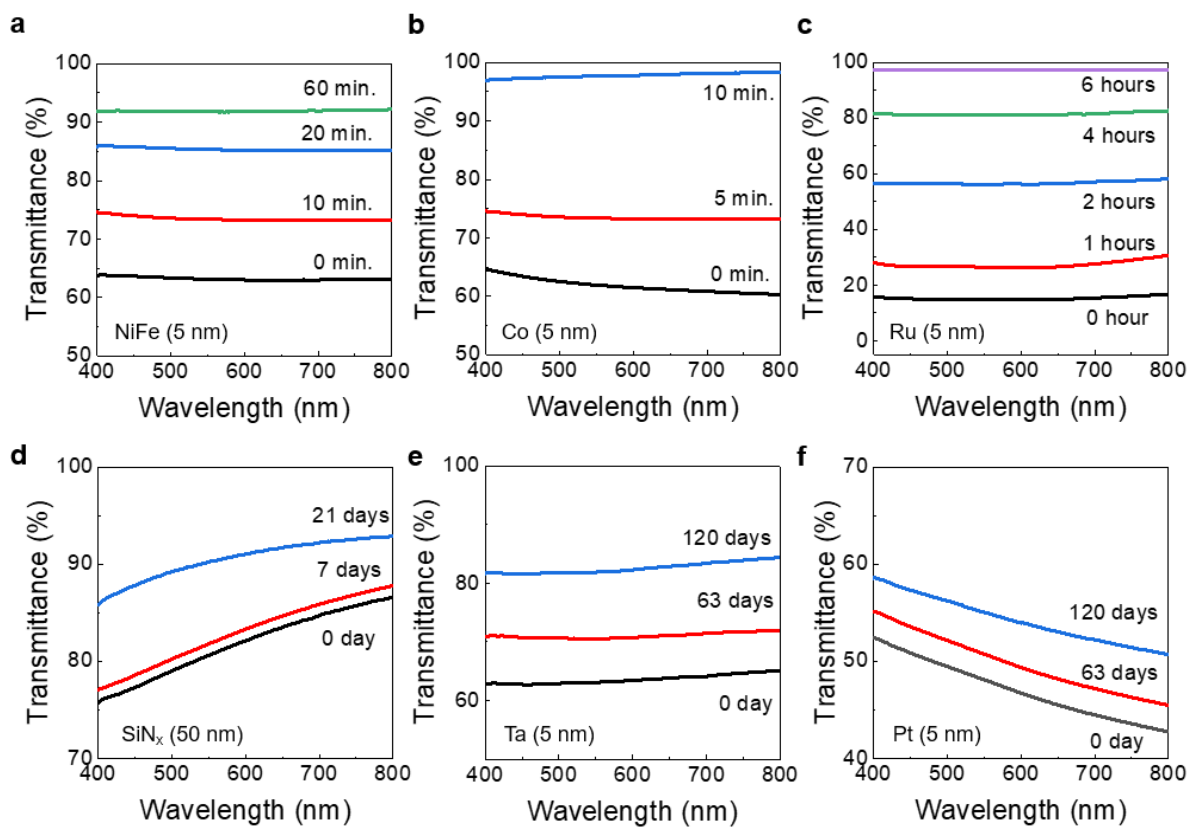

**Figure S3.** UV–Vis transmittance spectra of various MTJ-related thin films during dissolution in PBS solution. a) Transmittance change of a 5-nm-thick NiFe film, showing significant dissolution in PBS (pH 7.4) at 37 °C within 60 minutes. b) Transmittance behavior of a 5-nm-thick Co film, exhibiting rapid degradation within 10 minutes. c) Transmittance evolution of a 5-nm-thick Ru film, showing a pronounced increase over 6 hours, supporting its suitability as a bio-resorbable electrode. d) Transmittance of a 50-nm-thick $SiN_x$ layer, indicating gradual degradation over 21 days. e) Transmittance change of a 5-nm-thick Ta film over 120 days, showing negligible degradation. f) Transmittance change of a 5-nm-thick Pt film under the same conditions, also exhibiting poor bio-resorbability.

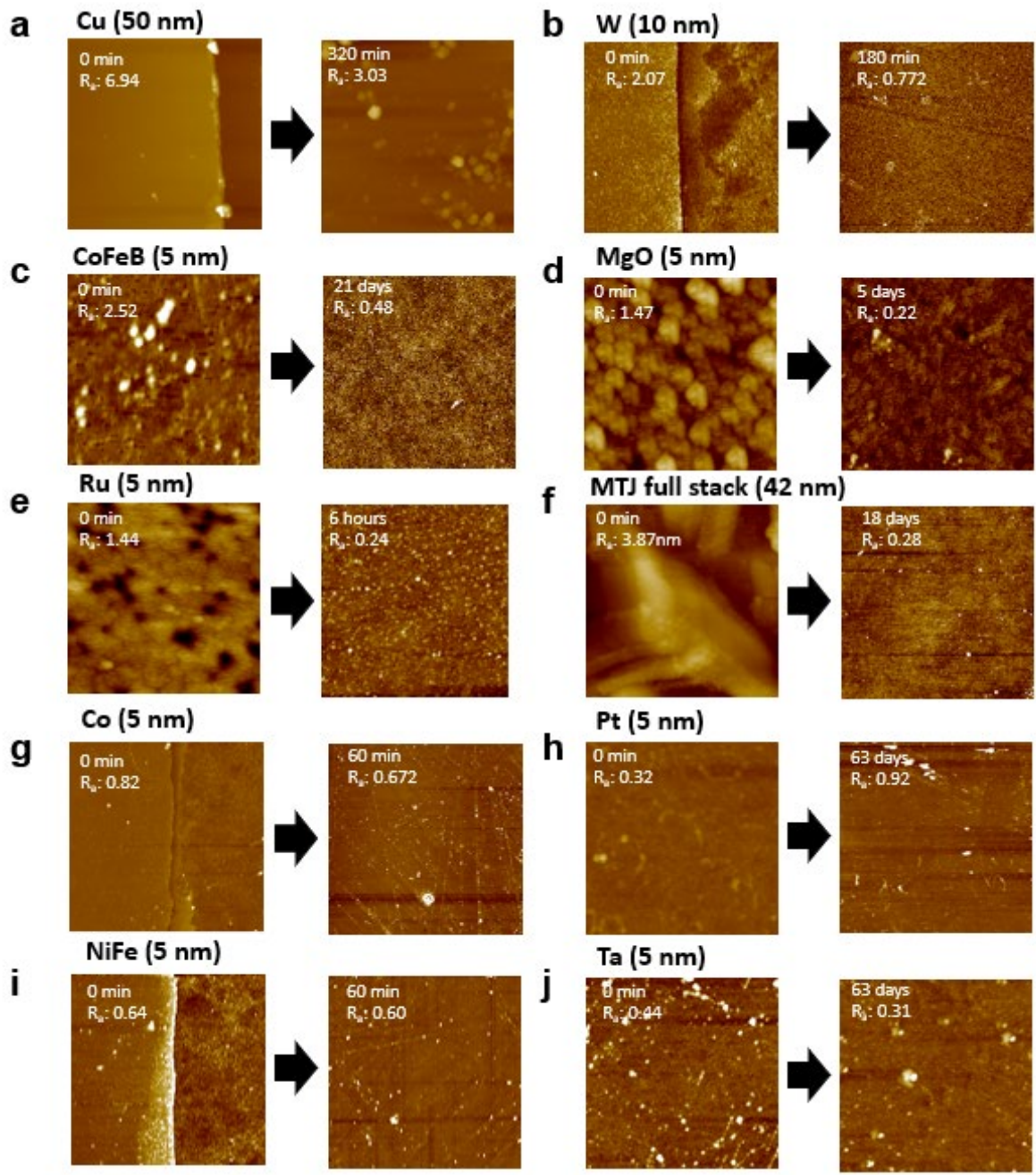

**Figure S4.** Atomic force microscopy (AFM) images showing surface roughness evolution of various MTJ-related materials during dissolution in PBS solution (pH 7.4). a–j) Each panel compares the AFM surface morphology of the film at the initial time point (left) and after dissolution (right). (a) Cu (50 nm), (b) W (10 nm), (c) CoFeB (5 nm) (d) MgO (5 nm), (e) Ru (5 nm), (f) MTJ full stack (42 nm), (g) Co (5 nm), (h) Pt (5 nm), (i) $Ni_{19}Fe_{81}$ (5 nm), and (j) Ta (5 nm).

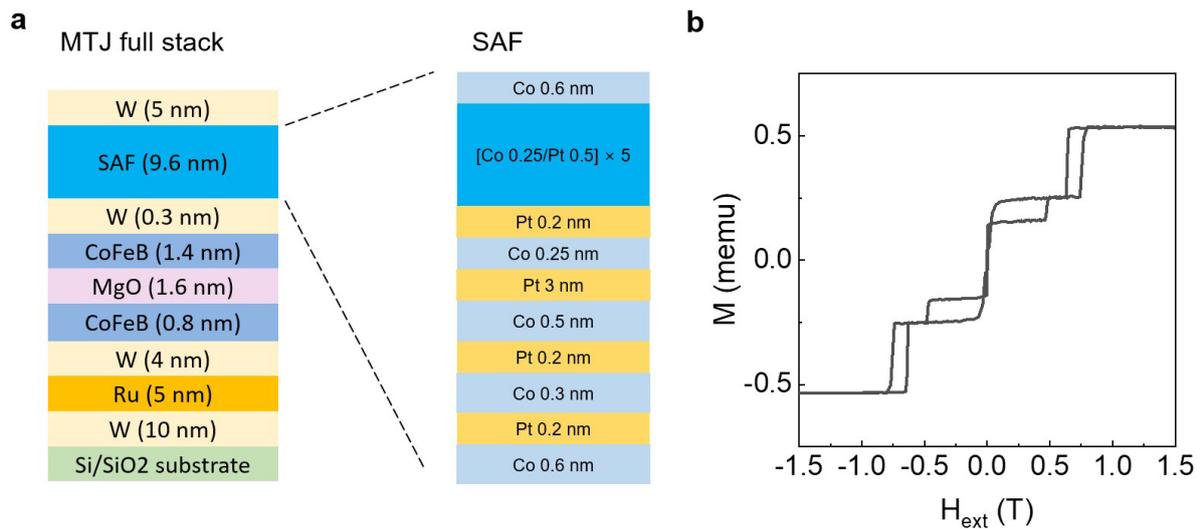

**Figure S5.** Structure and magnetic characterization of the MTJ full stack with the synthetic antiferromagnetic (SAF) layer. a) Schematic of the complete MTJ full stack including a 9.6-nm-thick Co/Pt-based SAF layer positioned above the CoFeB/MgO/CoFeB layers. The SAF structure is engineered to provide strong perpendicular magnetic anisotropy and synthetic antiferromagnetic coupling for thermal stability. b) The magnetization loop of the MTJ full stack was measured using vibrating sample magnetometry, exhibiting stepwise magnetic switching characteristics of SAF behavior and confirming the successful decoupling of the free and fixed magnetic layers within the MTJ structure.

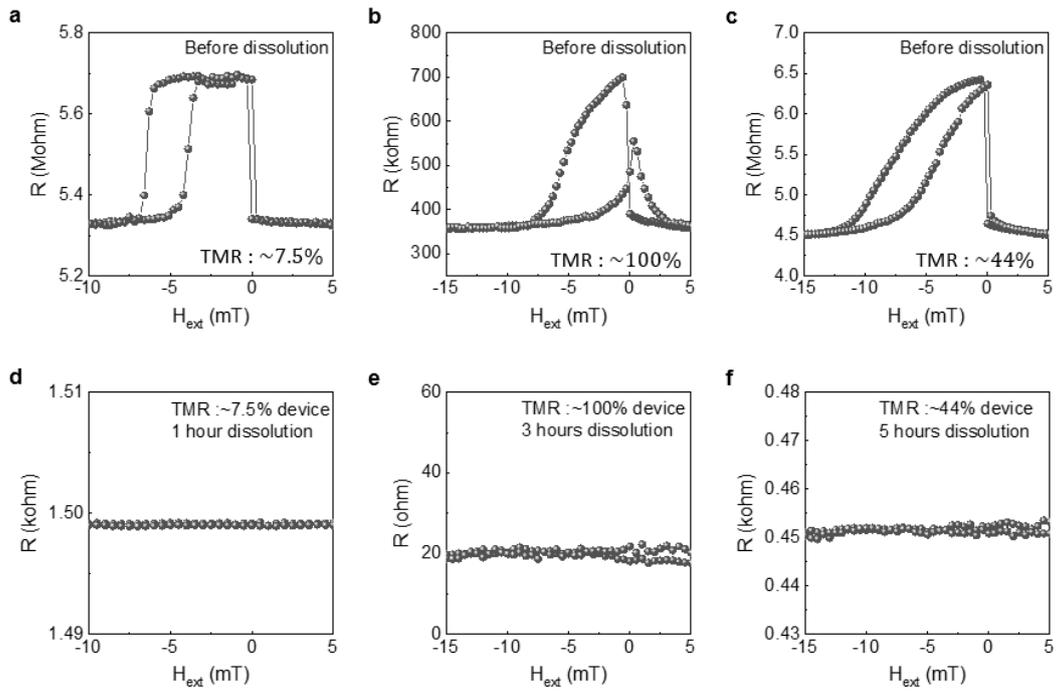

**Figure S6.** Dissolution behavior of the magnetic field ($H_{ext}$) driven TMR curve of various MTJ devices in PBS at 37 °C. a) TMR curve of a device with 7.5% TMR before dissolution. b) TMR curve of a device with 100% TMR before dissolution. c) TMR curve of a device with 44% TMR before dissolution. d) After 1 hour of dissolution of the 7.5% TMR device. e) After 3 hours of dissolution of the 100% TMR device. f) After 5 hours of dissolution of the 44% TMR device.